\documentclass[fleqn,usenatbib]{mnras}

\usepackage{newtxtext,newtxmath}
\usepackage{bm}
\usepackage{float}

\usepackage{graphicx}	
\usepackage{amsmath}	
\usepackage{multirow}

\newcommand{\Arctanh}{\rm Arctanh}
\newcommand{\Arcsin}{\rm Arcsin}
\newcommand{\arccoth}{\rm Arccoth}
\newcommand{\vc}{v_{\rm c}}
\newcommand{\ac}{a_{\rm c}}
\usepackage{amsmath,graphicx}
\newcommand{\wyn}[1]{{ #1}}

\usepackage{scalerel,tikz}
\usetikzlibrary{svg.path}
\definecolor{orcidlogocol}{HTML}{A6CE39}
\tikzset{orcidlogo/.pic={
 \fill[orcidlogocol] svg{M256,128c0,70.7-57.3,128-128,128C57.3,256,0,198.7,0,128C0,57.3,57.3,0,128,0C198.7,0,256,57.3,256,128z};
 \fill[white] svg{M86.3,186.2H70.9V79.1h15.4v48.4V186.2z}
 svg{M108.9,79.1h41.6c39.6,0,57,28.3,57,53.6c0,27.5-21.5,53.6-56.8,53.6h-41.8V79.1z M124.3,172.4h24.5c34.9,0,42.9-26.5,42.9-39.7c0-21.5-13.7-39.7-43.7-39.7h-23.7V172.4z}
 svg{M88.7,56.8c0,5.5-4.5,10.1-10.1,10.1c-5.6,0-10.1-4.6-10.1-10.1c0-5.6,4.5-10.1,10.1-10.1C84.2,46.7,88.7,51.3,88.7,56.8z};
}}
\newcommand\orcidicon[1]{\href{https://orcid.org/#1}{\mbox{\scalerel*{
\begin{tikzpicture}[yscale=-1,transform shape]
\pic{orcidlogo};
\end{tikzpicture}
}{|}}}}

\begin{document}

\title[Eccentricity Law]{The Eccentricity Distribution of Wide Binaries}

\author[Benisty et al]{~David Benisty~\orcidicon{0000-0002-9578-3081}$^{1,2}$,
N. Wyn Evans~\orcidicon{0000-0002-5981-7360}$^{3}$, Anne-Christine Davis$^{1,2}$
\\
$^{1}$ DAMTP, Centre for Mathematical Sciences, University of Cambridge, Wilberforce Road, Cambridge CB3 0WA, UK,\\
$^{2}$ Kavli Institute of Cosmology (KICC), University of Cambridge, Madingley Road, Cambridge, CB3 0HA, UK,\\
$^{3}$ Institute of Astronomy, University of Cambridge, Madingley Road, Cambridge CB3 0HA, UK.}

\maketitle
\begin{abstract}
Future {\it Gaia} and {\it Legacy Survey of Space and Time} data releases, together with wide area spectroscopic surveys, will deliver large samples of resolved binary stars with phase space coordinates, albeit with low-cadence. Given an eccentricity law $f(\epsilon)$, we derive properties of (i) the velocity distribution $v/\sqrt{G M/r}$ normalised by the value for a circular orbit at the measured separation $r$; (ii) the\wyn{astrometric} acceleration distributions $a/\left(G M/r^2\right)$ again normalised to the circular orbit value. Our formulation yields analytic predictions for the full statistical distribution for some commonly used eccentricity laws, if the timescale of data-sampling is comparable to or exceeds the binary period. In particular, the velocity distribution for the linear eccentricity law is {\it surprisingly simple}. With Bayesian analysis, we suggest a method to infer the eccentricity distribution based on the measured velocity distribution.
\end{abstract}


\section{Introduction}

\label{sec:intro}
Eccentricity is affected by a very wide range of physical processes. For example, tidal circularization and synchronization, mass loss through winds or mass gain through accretion, interactions with distant companions driving the Kozai effect, flybys or encounters with unbound objects in the local Galactic environment are all drivers of eccentricity changes. If the binaries have exchanged energy many times and reached statistical equilibrium, then their eccentricity distribution is thermalised like $f(e) = 2e$ ~\citep{Je19,Am37}. On the other hand, an excess of circular orbits is expected if the dominant effect is tidal circularization on the main sequence or pre-main sequence~\citep{Za77}. Given this variety, resolved binaries are not a homogeneous population and eccentricity laws need to be extracted at sub-population level~\citep[e.g.,][]{MS17}. 

For short period visual binaries, well-sampled orbital fits can provide all the elements. For longer period or wide binaries, this is not the case. Nonetheless, if low-cadence photometric and spectroscopic data are available for samples of wide binaries (e.g. from space missions like {\it Gaia} or ground-based telescopes like the {\it Legacy Survey of Space and Time}), then eccentricity laws can still be inferred.

This idea goes back to~\citet{To98}, who proposed to extract eccentricity laws using the distribution of the angle $\gamma$ between the radius vector that connects the two components and the vector of their relative motion. This was elaborated in \citet{To16}, who used both $\gamma$ and relative speed $v$ to infer eccentricity from the short observed arcs of orbits. \citet{To20} applied this method to the catalogue of \citet{ER18} to extract the eccentricity law, otherwise inaccessible owing to very long orbital periods. It was found that the eccentricity distribution is close to the thermal one, namely $f(e) = 2e$. Recently, \citet{HHZ} used simply the angle $\gamma$, applying a Bayesian approach to the sample of wide binaries in {\it Gaia} Early Data Release 3. They showed that the eccentricity distribution of wide binaries at separations of $\sim 100$ AU is close to uniform, but the distribution becomes super-thermal at separations greater than $1000$ AU.

There has also been much attention on using wide binaries to search for deviations from Newtonian gravity. This is because the two-body problem has a simple solution both in Newtonian gravity and in many modified gravity theories~\citep[e.g.,][]{ZBB,FLD,BD,Ben22}. Here, the focus is on the properties of the distribution of instantaneous velocities normalised to the circular velocity $v/\vc$. The study of these distributions was initiated by \cite{Pi18}, who pointed out that observationally there is a long tail that starts at $v/v_{\rm c} > 1 $ which may be possible evidence for the existence of modified gravity~\cite[see e.g.,][]{He12,Pi19,He19,1902.01857}. However, \citet{Cl20} argued that this tail is instead explicable in terms of a population of hidden triples, where one of the components of the wide binary is itself a close binary unresolved in the data.

The plan of the work is as follows. Section~\ref{sec:KepMot} sets out the background. Section~\ref{sec:ana} studies the statistical properties of the distributions of velocity and acceleration, and suggests new methods to infer the eccentricity distribution, while Section~\ref{sec:res} sums up.

\section{The Two Body Problem}
\label{sec:KepMot}
We begin by summarizing some facts about the two body problem. The ratio of 3-D relative velocity to the Newtonian circular velocity or $v/v_{\rm c}$ is
\begin{equation}
\frac{v}{\vc} = \frac{v}{\sqrt{G M /r}} = \sqrt{1 + e \, \eta},
\label{eq:vvcDef}
\end{equation}
which runs between $\sqrt{1\pm e}$ with frequency (or mean motion) $n$. Here, $e$ is the eccentricity, 
$G$ is the Newtonian constant and $M =M_1+M_2$ is the total mass of the two bodies and $r$ is the separation of the two components. 
Two running angles are used to describe the instantaneous position on the ellipse, namely $f$ or the true anomaly and $E$ or the eccentric anomaly, which we represent as $\eta = \cos{E}$ \citep[see e.g.,][]{Mu99}. \wyn{A change in radial velocity of a star over years can be measured by multi-epoch spectroscopy.  However, here we assume the acceleration is only measured astrometrically through a change in the proper motion over time, so only the acceleration component normal to the trajectory -- which produces the curvature -- is extractable.} The ratio of this \wyn{astrometric} acceleration $a$ to the Newtonian circular acceleration $a_{\rm c}$ \wyn{for the star of mass $M_1$} is
\begin{equation}
\frac{a}{a_{\rm c}} =  \frac{a}{G M/r^2} = {M_1\over M}\sqrt{\frac{1 - e^2}{1 - e^2 \eta^2}},
\label{eq:aacDef}
\end{equation}
which runs between $\sqrt{1-e^2}$ and 1 with frequency $2n$. 
%
%

\begin{figure}
\includegraphics[width=0.46\textwidth]{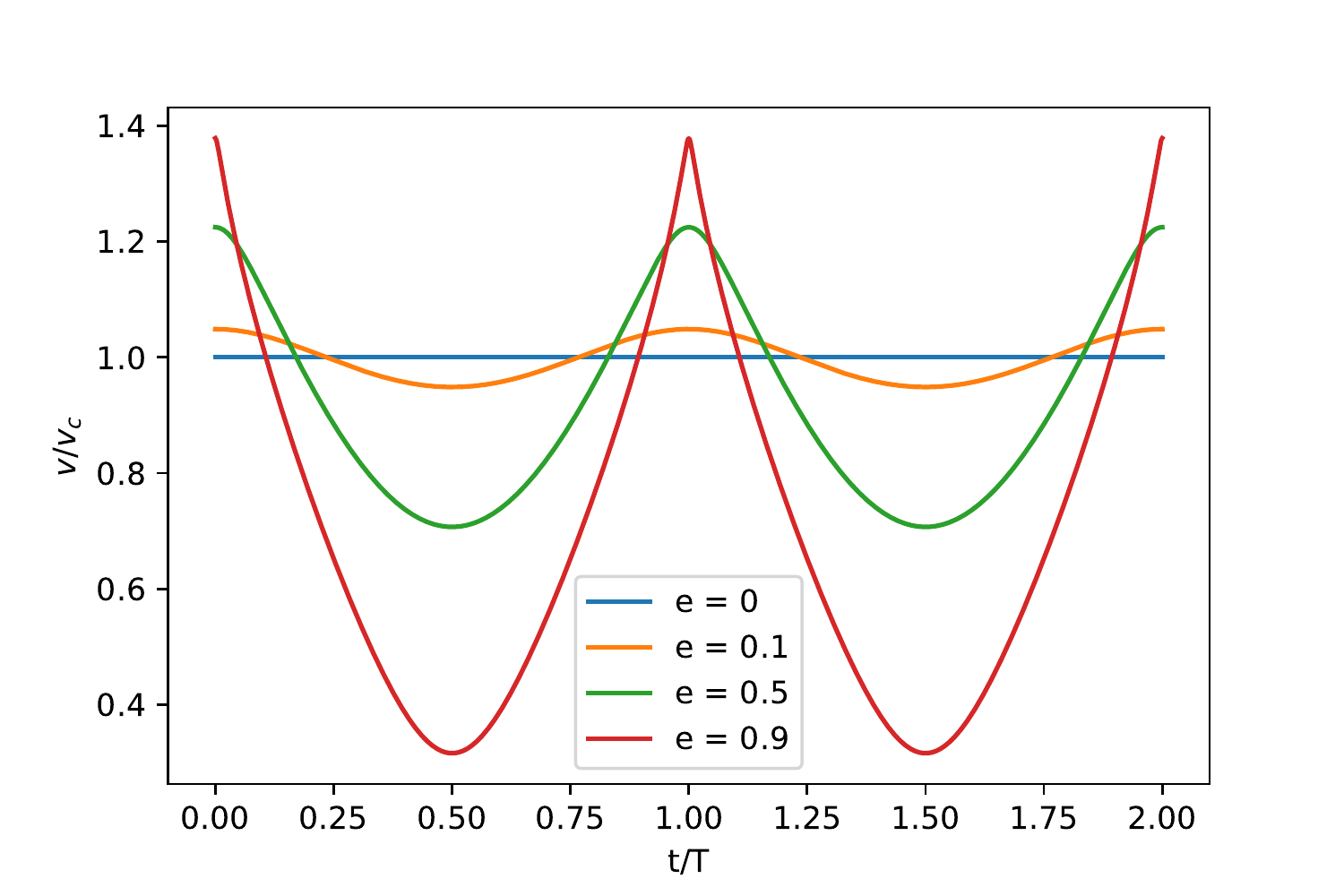}
\includegraphics[width=0.46\textwidth]{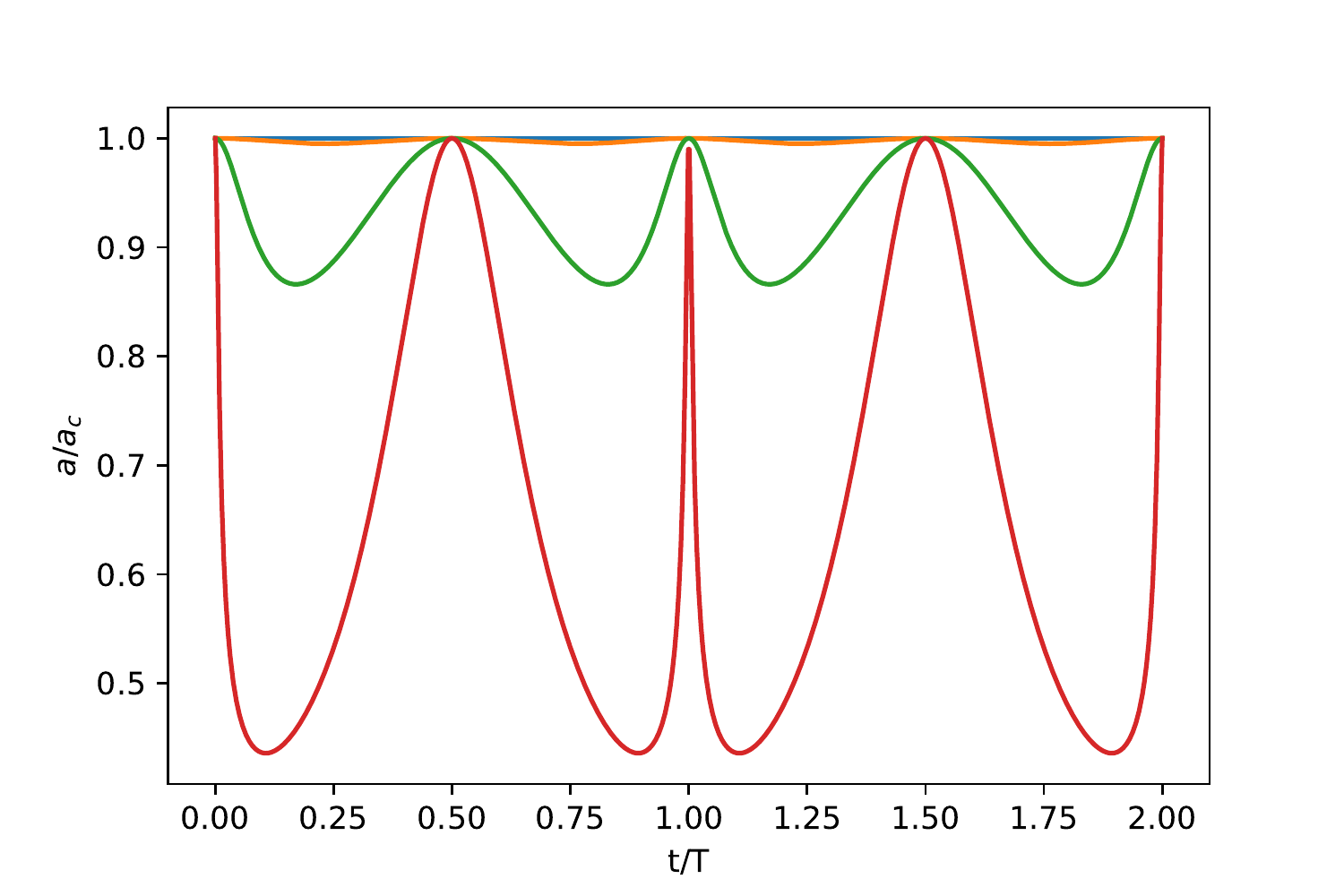}
\caption{The normalised velocity $v/\vc$ and accelerations $a/a_{\rm c}$ for different eccentricities values versus time in units of the period, $t/T$. The normalised velocity oscillates between $\sqrt{1 \pm e}$ with the frequency $n$. The normalised acceleration oscillates between $\sqrt{1 - e}$ and $1$, with the frequency $2n$. \wyn{(We have assumed that the masses of the two components satisfy $M_2\ll M_1$ so $M_1/M \approx 1$).}}
\label{fig:vvckeptime}
\end{figure} 
\begin{figure}
 	\centering
\includegraphics[width=0.49\textwidth]{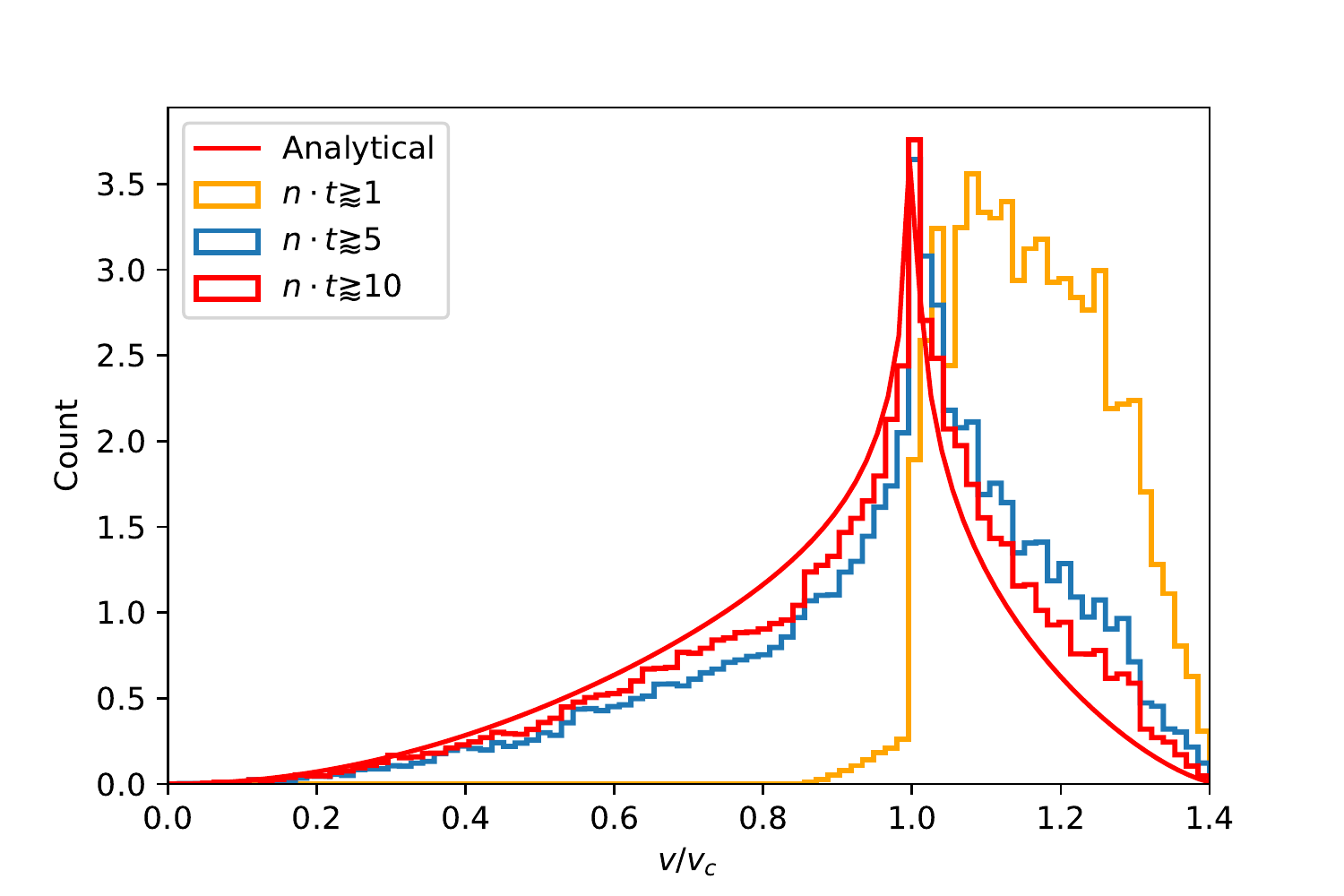}
\caption{{The distribution of $v/v_{\rm c}$ via simulations using direct orbital integrations. The simulations have a uniform prior of mean motion $n \in [0,n_{\rm max}]$ and eccentricity $e \in [0,1]$ with $10^3$ systems and $10^3$ samplings. The theoretical expectation as $n_{\rm max}t \rightarrow \infty$ is shown as a red curve (eq.~\ref{eq:mainlinres}).}} 
\label{fig:vvckepUniNF}
\end{figure} 

We study the case of a binary for which both masses and all six relative separation and velocity components are available. The masses of main sequence stars can be inferred using mass–luminosity relations (for example, equation 3 of \citet{Pi19} or Figure 1 of \citet{EvOh}). The parallaxes and proper motions of resolved binaries are available from {\it Gaia's} astrometry, whilst the line of sight velocity from cross-matches with spectroscopic surveys or from follow-up \citep[see][for a discussion of the difficulties]{Ba19}. The three-dimensional separation $r$ between the two components is hardest to measure. However, even if this is unavailable, we can still compute the circular velocity at the projected separation $R$. The ratio of $v$ to the circular velocity at the projected separation $\vc(R)$ is a good proxy for $v/\vc(r)$ \citep{Pi18}. The two are related by $\sqrt{\sin\theta}$, where $\theta$ is the unknown angle between the current binary separation vector and the line-of-sight. This can be computed for random alignments and its mean value is close to unity.  \wyn{So, though the three-dimensional separation of any binary is almost impossible to measure even with {\it Gaia}, the sky-projected separation is readily measurable and can be combined with statistics on the three-dimensional to two-dimensional projection factors.}
 
Though more difficult, the distribution of $a/a_{\rm c}$ may also become accessible. If the binary period is larger than {\it Gaia}'s baseline, the orbital motion of a binary appears approximately linear. However, by comparing the measured proper motion of a source over a sufficiently long time period, a change in proper motion over time can be detected~\citep{Ke19}. \cite{Br18} compared proper motions measured by {\it HIPPARCOS} and {\it Gaia}, producing a catalog of astrometrically accelerating stars. 

Fig.~\ref{fig:vvckeptime} shows the dependence of $v/\vc$ and $a/a_{\rm c}$ vs. $2\pi t/T$. It is easy to show that $v/\vc$ oscillates between $\sqrt{1 \pm e}$ with the frequency $n$. For $v/\vc = a/a_{\rm c} = 1$, the curve describes a circular motion, otherwise the motion is elliptical (for bound orbits). The acceleration ratio $a/a_{\rm c}$ equals to the ratio $\left(r/\rho\right)^2$. For a circular motion, the radius of curvature $\rho$ is identical to the separation $r$. For any eccentricity $e>0$ the radius of curvature $\rho$ is larger then the separations $r$. Therefore, for $e>0$ the ratio is smaller then one $a/a_{\rm c} < 1$, as the lower panel of Fig~\ref{fig:vvckeptime} shows.

\section{A Statistical Theorem}
\label{sec:ana}

The probability distributions are analytic when the sampling time $t_s$ exceeds the typical periodic time of the systems $T = \sqrt{a_{sa}^3/G M} \ll t_s$. Let us suppose the eccentricity law is $f(e)$, normalised to unity. Then, by differentiation of the mean anomaly equation, the joint probability of measuring a binary with eccentricity $e$ and eccentric anomaly $E$ is 
\begin{equation}
P(e,E) \, de dE =
{\left(1-e \cos E\right) f(e) \,de\,dE \over \int_{t} n dt}.
\label{eq:secInt}
\end{equation}
Here, we have assumed that the system has performed enough periods so we can take the maximal mean anomaly to $\infty$. This condition leads to the approximation $ n\,dt \approx dE$ for large sampling time, since $e \sin E \ll E$. It is possible to simplify the integration over the anomaly via  $\eta = \cos E$. The joint probability becomes:
\begin{equation}
P(e,\eta) \, de d\eta = {1\over \pi}
\frac{1-e \, \eta} {\sqrt{1-\eta^2}} f(e) \,de\, d\eta.
\label{eq:jointpdf}
\end{equation}
%
%

It follows that if $\hat{A}\left(e,\eta\right)$ is a parameter that depends directly on the eccentricity and the anomaly, then the mean of the parameter is given by the integral:  
\begin{equation}
\Bigl\langle{A}\Bigr\rangle = \frac{1}{\pi}\int_{-1}^{1} \int_{e} \hat{A}\left(e,\eta\right)\,\frac{1- e \eta }{\sqrt{1-\eta^2}}f(e) \, d\eta \, de.
\label{eq:theorem}
\end{equation}

From eq.~(\ref{eq:jointpdf}), we can compute the marginalised distribution of normalized velocities $X=v/\vc$ by transforming $(\eta,e) \rightarrow (X,e)$ to obtain
\begin{equation}
P(X)dX = \frac{2}{\pi}X (2-X^2) dX \int_0^1 {f(e)\,de\over \sqrt{e^2 - (X^2-1)^2}}
\end{equation}
Similarly, the marginalised distribution of normalized accelerations $Y = a/\ac$ is
\begin{equation}
P(Y)dY = {1\over \pi} {Y dY\over \sqrt{1-Y^2}} \int_0^1 f(e)\, de \left[ {Y\over\sqrt{Y^2- (1-e^2)}} -1\right].
\end{equation}
%
%

%
\begin{figure}
 	\centering
\includegraphics[width=0.49\textwidth]{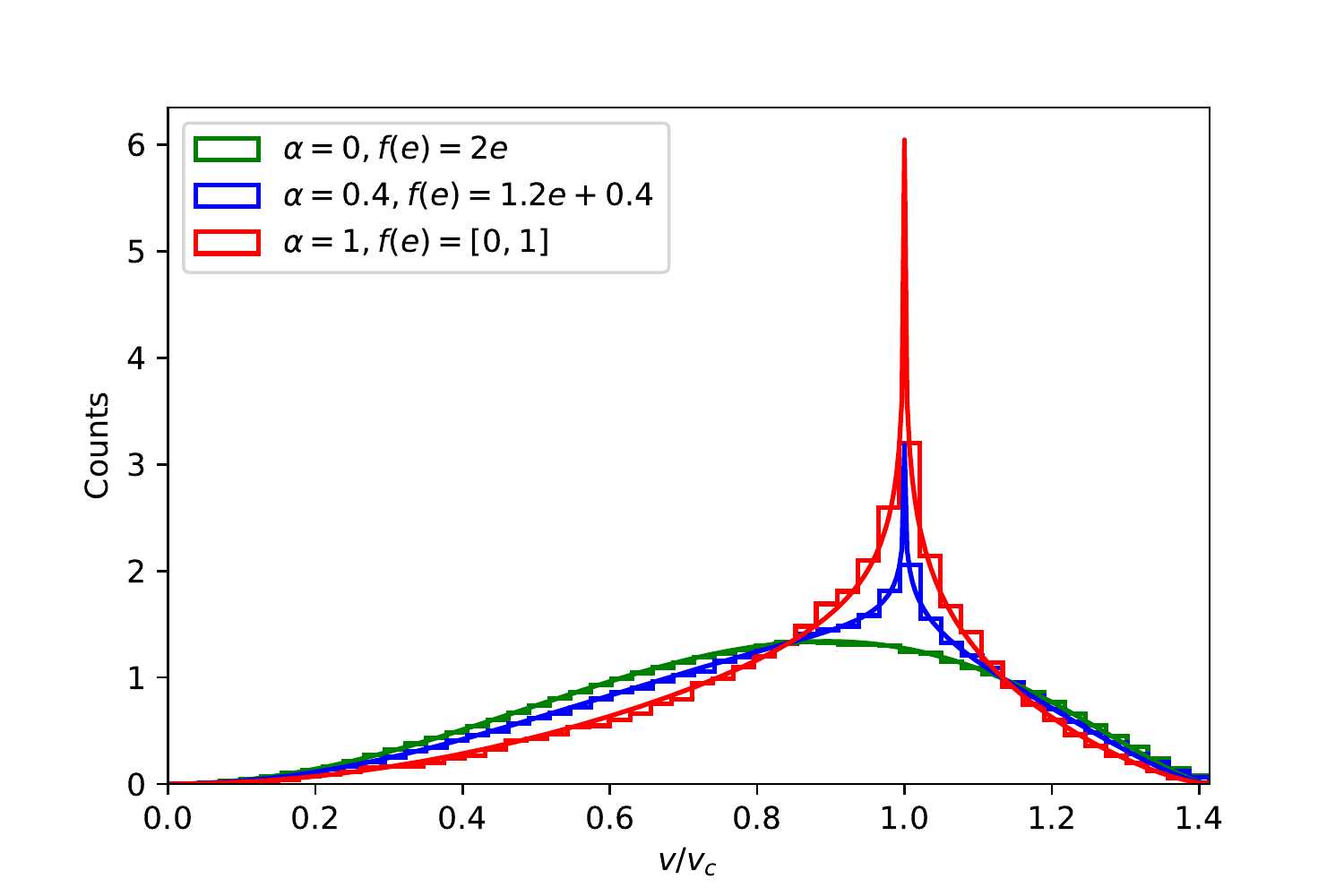}
\\
\includegraphics[width=0.49\textwidth]{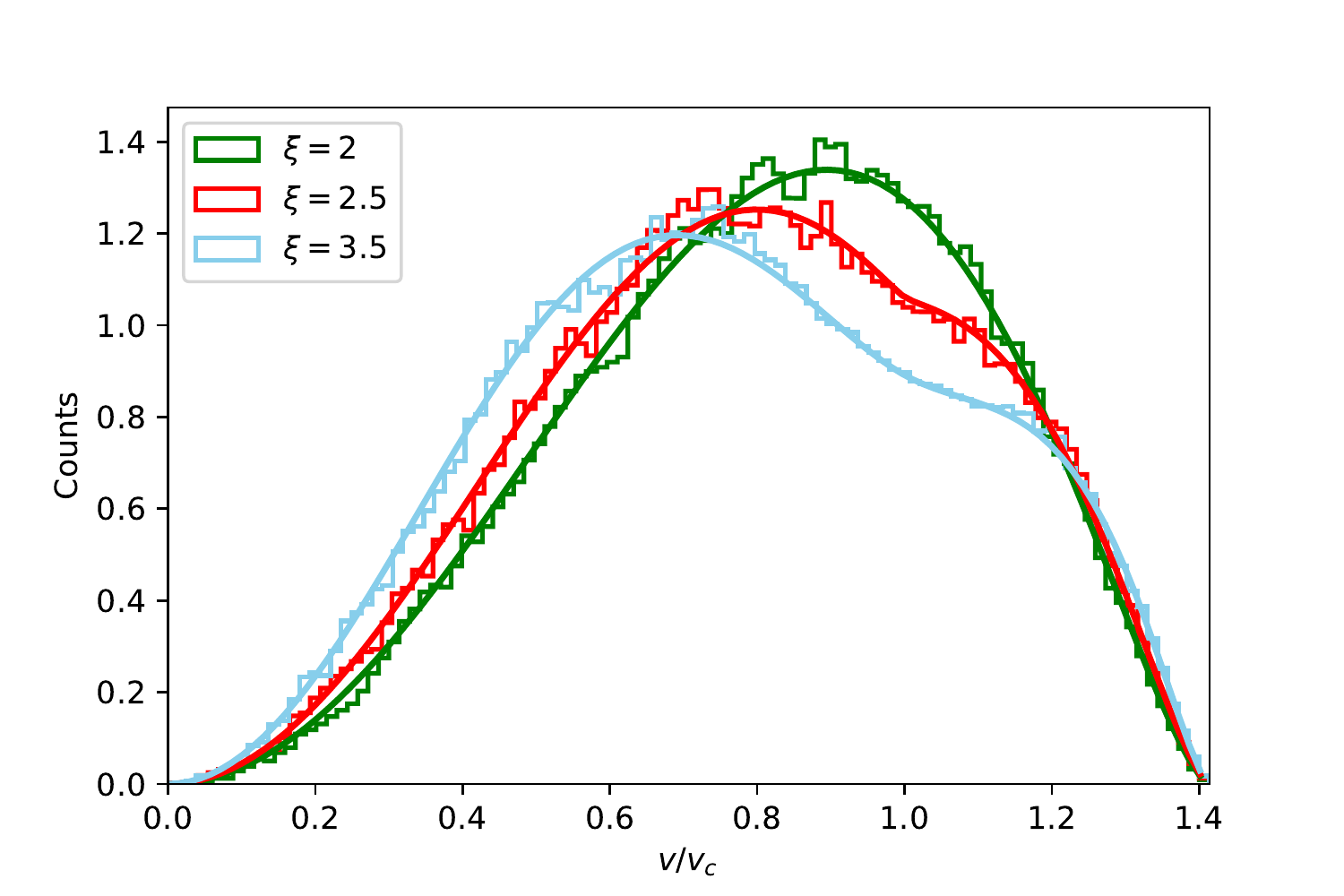}

\caption{{Upper: The simulations for the $v/\vc$ for the linear eccentricity law with $\alpha =0$ (thermal, green), 0.4 \citep[][blue]{To16}, 1 (uniform, red) with uniform priors on the semimajor axis and on the masses ($M \in [0.,10] M_{\odot}$). The sampling time is three years. The analytical expressions (eq.~\ref{eq:mainlinres}) are shown as full curves, and the simulated distribution as histograms. Lower: The same, but for the power-law eccentricity distribution with $\xi = 2$ (thermal, green), $\xi =2.5$ (red) and $\xi = 3.5$ (blue). The full curves are the analytic expressions in  eq.~(\ref{eq:mainpowres}).}}
\label{fig:EccDisLinear}
\end{figure} 

\subsection{Linear distribution}
For simplicity, we take the eccentricity prior:
\begin{equation}
    f(e) = \alpha + 2(1-\alpha) e,
    \label{eq:linearecc}
\end{equation}
where $\alpha$ is a constants. Of course if $\alpha=1$, this is a uniform prior. If $\alpha=0$, this is the thermal prior. It also includes the eccentricity distribution of wide binary systems with $f(e) = 1.2 e + 0.4$ suggested by \cite{To16}. From Eq~(\ref{eq:jointpdf}) with the eccentricity distribution~(\ref{eq:linearecc}), the distribution of normalised velocities is:
\begin{equation}
P(X) = \frac{2 \Lambda^2}{\pi X} \left(2 (1\!-\!\alpha) \Lambda +\alpha  \Arctanh \Lambda  \right)
\label{eq:mainlinres}
\end{equation}
where $\Lambda = X\sqrt{2- X^2}$. {\it This is an astonishingly simple final result!} 

Note that as $X \rightarrow 0$, $P(X) \propto X^2$ and so grows quadratically, whilst as $X \rightarrow \sqrt{2}$, then $ P(X) \propto (\sqrt{2}-X)^{3/2}$. Irrespective of $\alpha$, the distributions pass through the same two points (which are the solutions of the transcendental equation $\tanh 2\Lambda = \Lambda$). 

The distribution of normalised accelerations is 
\begin{eqnarray}
        \label{eq:normacc}
    P(Y) &=& \frac{1}{\left(\alpha\!+\!3\pi\!-\!4\right) Y^2 \sqrt{1\!-\!Y^2}}\left[
        6 \alpha  Y \left(K(Y^2)\!-\!E(Y^2)\right) -\right.\nonumber \\
        & & \left. (3 \pi\!-\!4) (\alpha\!-\!1) Y^3\!+\!3 \alpha Y \sqrt{1\!-\!Y^2}-3 \alpha  \Arcsin Y \right]
\end{eqnarray}
As $Y \rightarrow 0$, $P(Y) \propto Y$, whilst as $Y \rightarrow 1$, $P(Y) \propto \log(1-Y)/(1-Y)^{1/2}$. We check the validity of condition ($T \ll t_{s}$) with simulations of orbit integrations of $10^3$ binaries. The simulations have a uniform prior of mean motion $n \in [0,n_{\rm max}]$ and uniform eccentricity ($\alpha =1$) with $10^3$ samplings, which is taken to be one year. The mean anomaly depends on the time and frequency only through the combination $n\dot t$ so taking different time samplings yields only a re-scaling of the parameters. For larger values of $n_{\rm max}$, the simulation samples more periods and the approximation approaches to the red curve of the integration. A condition for the validity of our theorem is that $n_{\rm max}t \gtrsim 5$, or that the binary period is comparable to the sampling time.

\begin{figure}
\centering
\includegraphics[width=0.43\textwidth]{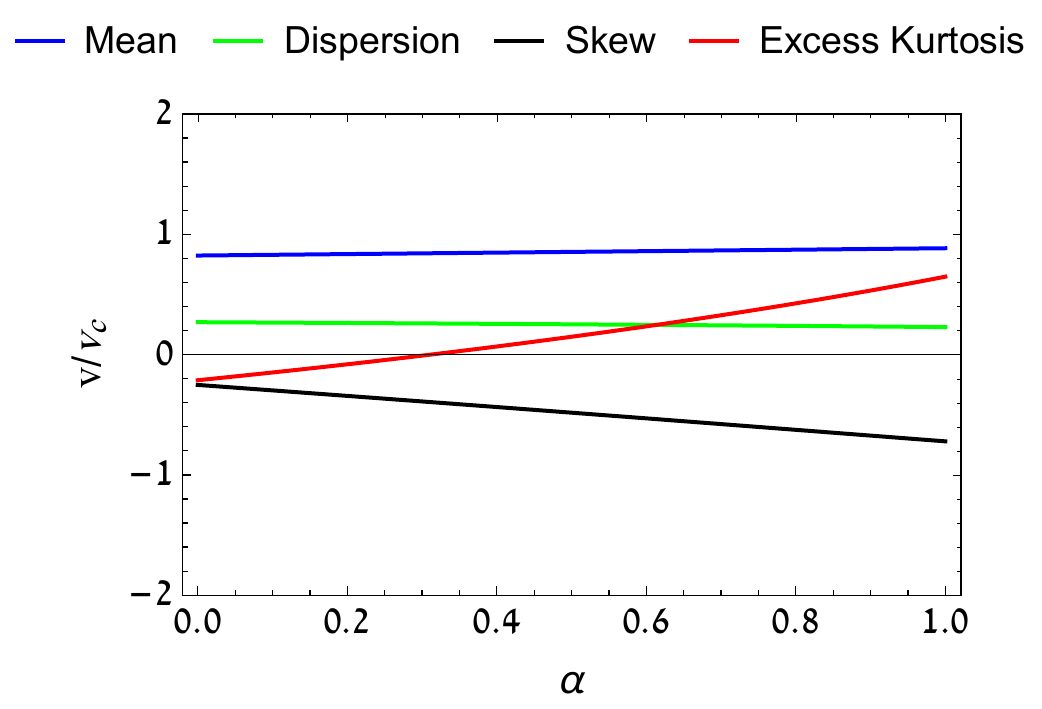}
\\
\includegraphics[width=0.36\textwidth]{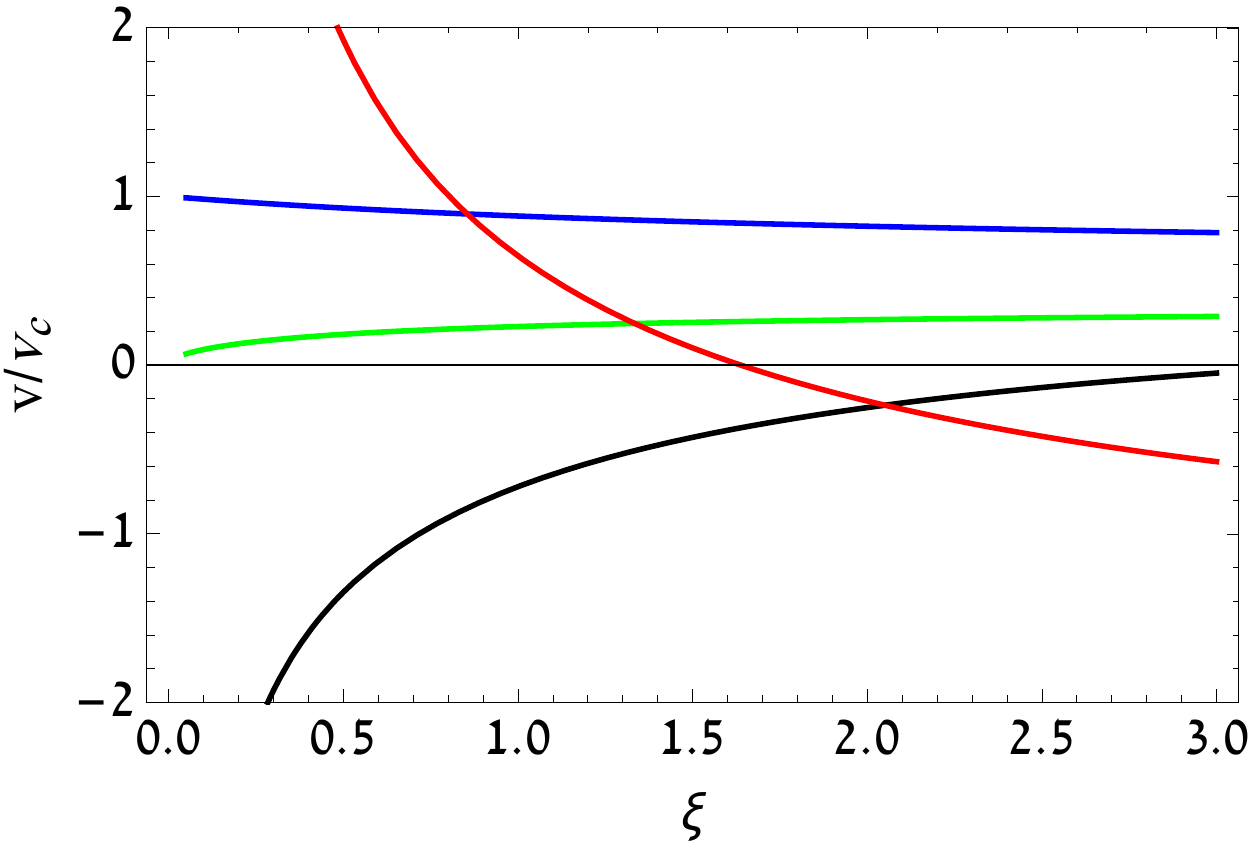}
\label{fig:DirectEccenMethod}
\caption{Mean, dispersion, skewness and excess kurtosis of the normalised velocity distribution $v/v_{\rm c}$ for (upper) the linear eccentricity distribution (\ref{eq:linearecc}) and (lower) the power-law (\ref{eq:powerlawec}). The fraction of eccentric orbits is increasing as $\alpha\rightarrow 1$ or as $\xi\rightarrow 0$. Notice the mean and dispersion are rather insensitive to the eccentricity law, but the skewness and excess kurtosis are not. The shape of the distribution changes considerably as more eccentric orbits are included, and this is encoded in the higher moments.}
\label{fig:kurt}
\end{figure} 

The upper panel of Fig~\ref{fig:EccDisLinear} shows the simulation results for the $v/\vc$ distribution against the theoretical prediction. There is excellent agreement. The mode is at $v/v_{\rm c} = 1$ for a uniform distribution ($\alpha =1$). Increasing the contribution of eccentric orbits causes the mode to move to lower values of $v/v_{\rm c}$, but also increases the heaviness of the tails. 

Notice that although the shape of the distribution changes substantially as $\alpha$ varies, the mean and variance do not. In fact, these quantities are analytic and have only a weak dependence on $\alpha$, namely
\begin{eqnarray}
\Bigl \langle {v \over v_{\rm c} } \Bigr \rangle &=&{{192\sqrt{2}\!+\!4\alpha(49\,\arccoth\sqrt{2}\!-\!27\sqrt{2})}\over 105 \pi},\nonumber\\
\text{Var}\left(\frac{v}{\vc}\right)
&=& {3\!+\!\alpha\over 4}\!-\!\Bigl \langle {v \over v_{\rm c} } \Bigr \rangle^2.
\end{eqnarray}
The behaviour of the moments \wyn{(inferred from numerical integrations)} is illustrated in the upper panel of Fig.~\ref{fig:kurt}. The distributions are negatively skewed as the left tail is longer and heavier than the sharply truncated right tail. This is consistent with the mean being less than the median for a unimodal distribution. As the contribution of eccentric orbits increases ($\alpha \rightarrow 0$), the distribution becomes less skewed, as the right and left tails become more equal. The excess kurtosis is positive (has fatter tails than a Gaussian) for a uniform eccentricity distribution ($\alpha =1$), but becomes negative as $\alpha \rightarrow 0$ (thinner tails).

\begin{figure}
 	\centering
\includegraphics[width=0.49\textwidth]{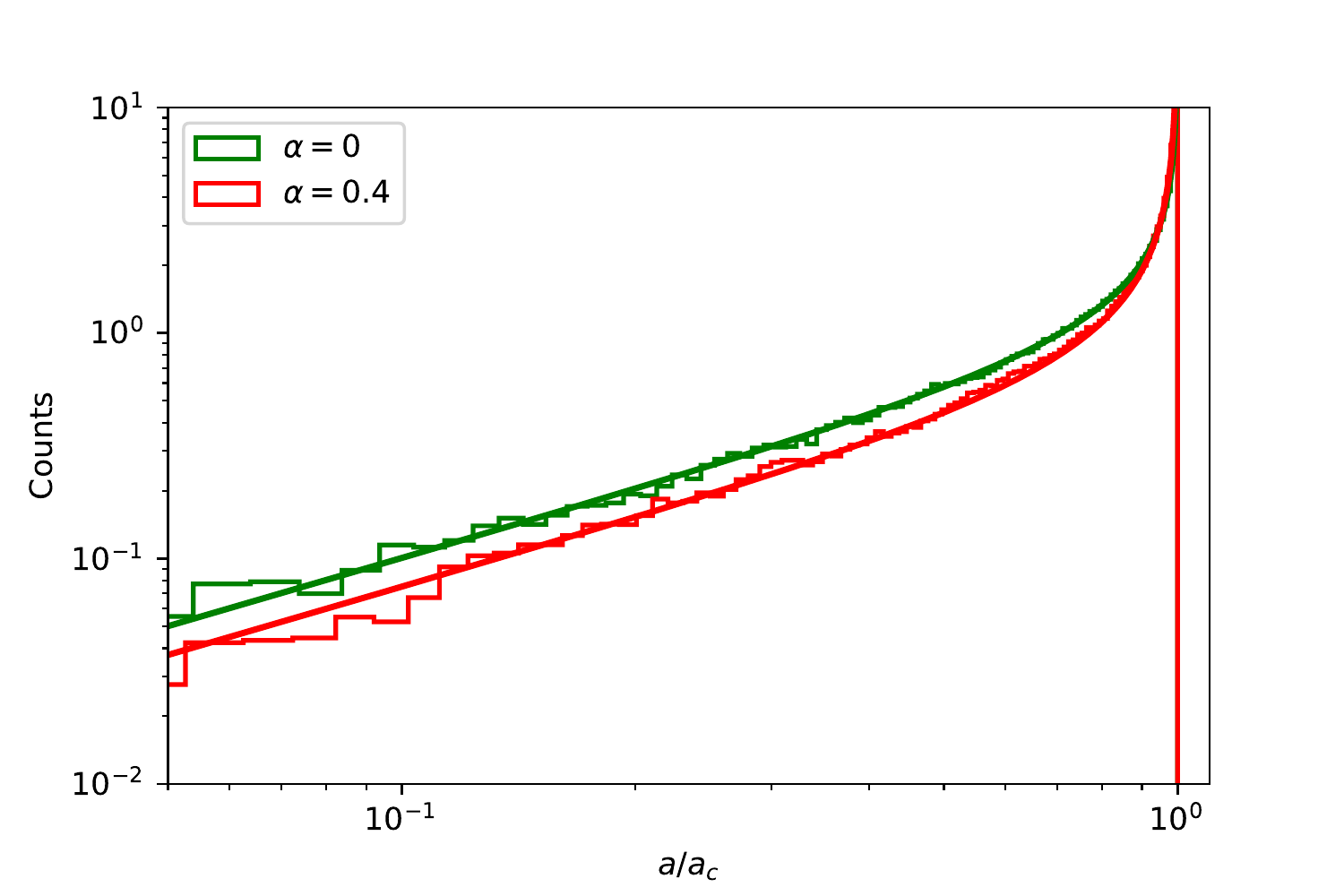}
\caption{{The simulations of the normalised accelerations $a/a_c$ for $\alpha =0$ (thermal, green), 0.4 \citep[][red]{To16}, with uniform priors on mass and semimajor axis. The sampling time is three years. The analytical expressions (eq.~\ref{eq:normacc}) are shown as full curves, and the simulated distribution as histograms.}}
\label{fig:aac}
\end{figure} 

Finally, we also show in Figure~\ref{fig:aac} a comparison between the normalised acceleration distributions $Y$ from the simulations against the theoretical result in eq~(\ref{eq:normacc}). The normalised acceleration $Y= a/a_{\rm c}$ for low eccentricities is around one, since the radius of curvature is close to the separation. However, thermal  distributions give a stronger preference for large eccentricities, then the radius of curvature is much larger then the separation. This is consistent with Figure~\ref{fig:aac} which shows the mode to be unity in all cases, but heavier tails when more eccentric orbits are present.

\subsection{Power-Law Distribution}

Another commonly used eccentricity distribution is the power-law \citep[cf][]{MS17}
\begin{equation}
f(e,\xi) = \xi \, e^{\xi-1},
\label{eq:powerlawec}
\end{equation}
where $0 \leq e \leq 1$. The distribution is normalized to one. Note that the case $\xi =2$ is the thermal distribution, whilst $\xi =1$ is the uniform prior. As $\xi \rightarrow 0$, eccentric orbits become increasingly disfavoured. Figure 36 of \citet{MS17} shows the power-law index of the eccentricity distribution for early-type and late-type binaries with $\xi$ typically in the range $0.75$ to $2$. Solar-type binaries have $\xi =1.5$, about halfway between uniform ($\xi =1$) and thermal ($\xi=2$) distributions. 

The corresponding quantities for the power-law distribution are also analytic, though a little more cumbersome. Here, we quote only the distribution of normalised velocities:
\begin{eqnarray}
P(X) &=& X (X^2-2) \Gamma \left( \frac{1 - \xi}{2} \right)
\times 
\Biggl[ \frac{2\sqrt{\pi} \left| X^2 - 1\right|^{\xi -1} } { \Gamma\left( -\frac{\xi}{2}\right) }  \nonumber \\
&+& \xi\, {}_2F_1\left(\frac{1}{2},\frac{1-\xi }{2};\frac{3-\xi }{2};\left( X^2 - 1\right)^2\right) \Biggr].
\label{eq:mainpowres}
\end{eqnarray}
The distribution is continuous, but its gradient is discontinuous at $X=1$ unless $\xi$ is an integer. The lower panels of Figure~\ref{fig:EccDisLinear} shows the theoretical distributions for different values of $\xi$, together with data from simulated binary orbits. We note that the mean of the distribution is again relatively insensitive to the power-law index $\xi$, at least within the physical range ($0 \le \xi \lesssim 3)$. The lower panel of Figure~\ref{fig:kurt} shows the moments. Again, we see that there is more discrimination between models in the higher moments, particularly the skew and excess kurtosis, as displayed in the right panel of Fig.~\ref{fig:kurt}. Now, the proportion of eccentric orbits increases as $\xi \rightarrow 0$. So, there is the same picture of negative skewness and positive excess kurtosis as the eccentric orbits begin to dominate over the circular ones.

Although we have only looked at two special cases for the eccentricity distribution, we have uncovered some interesting properties. The insensitivity of the mean of the eccentricities is valuable as it provides us with a means of measuring contamination from interlopers such as hierarchical triples in our samples. The sensitivity of the higher moments to eccentric orbits suggests that this is a powerful discriminant between the eccentricity laws themselves. 

\begin{figure}
 	\centering
\includegraphics[width=0.45\textwidth]{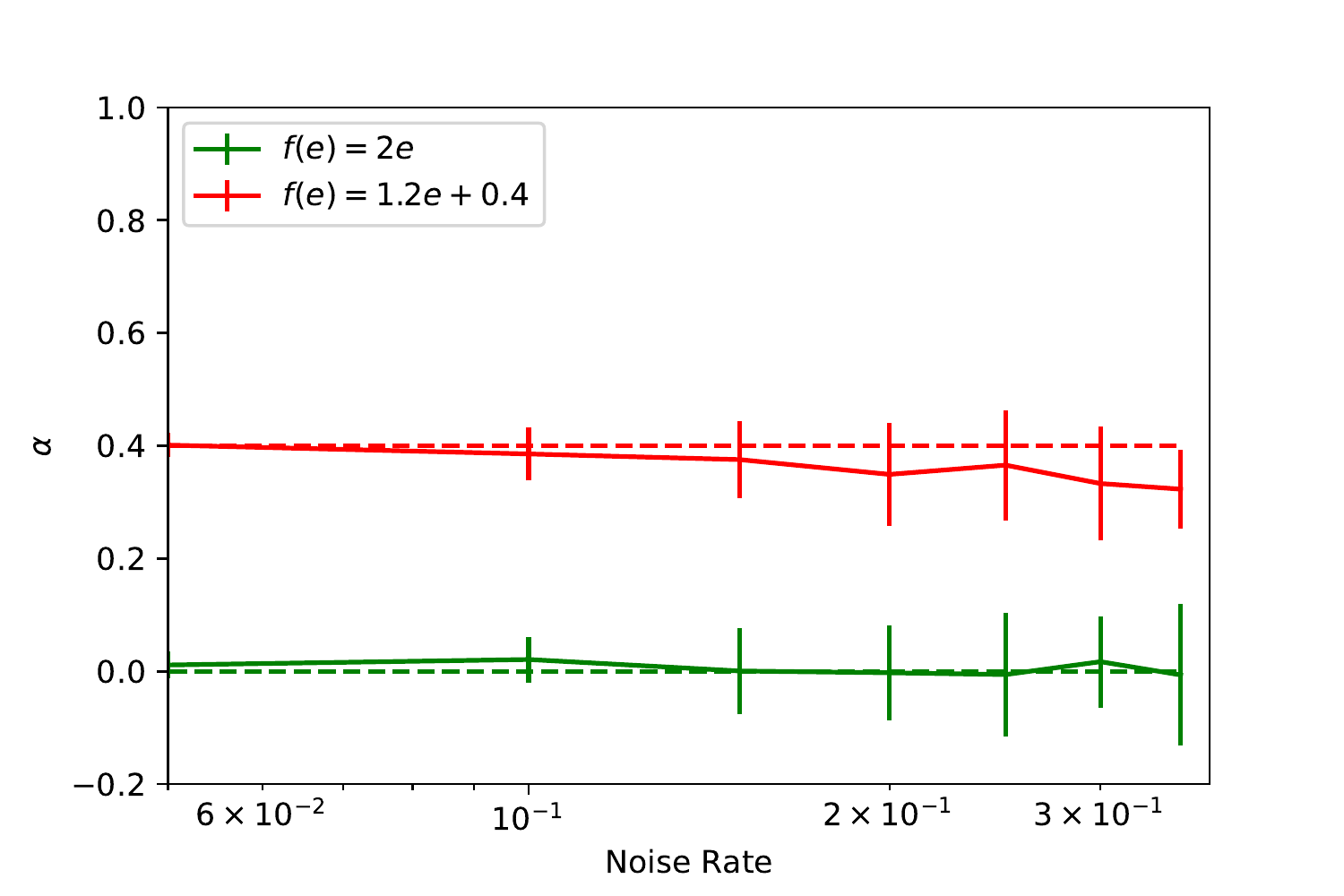}

\caption{The posterior of $\alpha$ versus different noise rate $\sigma_X$ for two examples of the linear eccentricity laws using eq~(\ref{eq:chisq}). The below table gives the models and the best fit for $\sigma_X = 0.05$ and $0.10$.}

\begin{tabular}{|c|c|c} 
  \hline\hline
     prior distribution & \multispan2{\hfill posterior value\hfill} \\
     \null& $\sigma_X=0.05$ & $\sigma_X=0.10$ \\
           \hline \hline
          $\alpha =0$ \\ (or $f(e)= 2e$) & $\left( 1.11 \pm 2.35 \right) \cdot 10^{-2}$ & $\left( 2.085 \pm 4.051 \right) \cdot 10^{-2} $ \\
             \hline 
            $\alpha =0.4$ \\ (or $f(e) = 1.2 e + 0.4$) &  $0.401 \pm 0.021$ & $0.385 \pm 0.047$ \\
                 \hline  \hline 
\end{tabular}
 	\label{fig:BayesFitLinear}
\end{figure} 

\subsection{Inference on the eccentricity distribution}
\label{sec:meth}

\wyn{Given an eccentricity law, we generate mock data for the $X= v/\vc$ distributions, adding Gaussian noise. Table~2 of \citet{Pi22} suggests error in $X$ of order $\sigma_X = 0.1-0.2$ is appropriate for the latest {\it Gaia} data releases. We added Gaussian noise $N(1,\sigma_X/\sqrt{2})$ to each of the two velocity components. As we have the exact distribution $F(X)$ in (\ref{eq:mainlinres}), we compute its convolution with a Gaussian $F(X,\sigma_X)$ and hence obtain $\sigma_P(X)= F(X,\sigma_X) - F(X)$.} 

Given the mock data $\{X^{(i)}\}$, we can construct the likelihood from the $\chi^2$, which is difference between the distribution of the mock data $F(X^{(i)})$ and the exact distribution $P(X^{(i)},\alpha)$ given in eq~(\ref{eq:mainlinres}), normalised by the uncertainty ${\sigma_P^{i}}$
\begin{equation}
\chi^2 = \sum_{i} \left(\frac{F(X^{(i)}) - P\left(X,\alpha\right) }{\sigma_{P}^{i}}\right)^2
\label{eq:chisq}
\end{equation}
Fig~\ref{fig:BayesFitLinear} shows different posterior values of $\alpha$ with different noise rates $\sigma_X$. For $\sigma_P\lesssim 5\%$, the mean with the error encompasses the initial value of $\alpha$. However, with a larger noise rate, the posterior does not fit for the prior value. This suggests that existing and future {\it Gaia} data releases may provide strong constraints on the eccentricity law from the normalised velocities, complementing alternative approaches~\citep[e.g.,][]{To98,To16,HHZ} 

\section{Conclusions}
\label{sec:res}

This paper derives formulae for the normalized velocity distributions $v/\vc$ and acceleration distributions $a/\ac$ for binary stars, as a function of the eccentricity law. Here, $\vc$ and $\ac$ are the velocity and acceleration of a circular orbit at the measured separation $r$. This requires knowledge of the masses of the two components of the binaries, which can be inferred from mass-magnitude relationships.

The study of these distributions was initiated by \cite{Pi18}, who pointed out their importance for tests of alternative theories of gravity. In the data, there is a long tail that starts at $v/v_{\rm c} > 1 $ which may be possible evidence for the existence of modified gravity~\cite[see e.g.,][]{He12,Pi19,He19}. However, \citet{Cl20} argued that this tail is instead explicable in terms of a population of hidden triples, where one of the components of the wide binary is itself a close binary unresolved in the {\it Gaia} data.

The normalized velocity and acceleration distributions also depend on the underlying eccentricity law. This fundamental property of binary star populations has only sparse constraints from data~\citep{MS17, To20, HHZ}. Provided the binary period is comparable to the sampling time, we derive a statistical theorem that allows us to calculate the properties of the distributions analytically. 

We study two eccentricity distributions in some detail, namely a linear law $f(e) = \alpha + 2(1-\alpha) e $ \citep[c.f.,][]{To16}, and a power law $f(e) = \xi e^{\xi-1} $. We derive the exact normalized velocity and acceleration distributions for these eccentricity laws. These are the principal results of this {\it Letter}. In particular, the velocity distribution for the linear law given in eq~(\ref{eq:mainlinres}) is surprisingly simple.
Simulations allow us to both verify our theorem and estimate the likely correction in the case of incomplete sampling. The means and variances of the normalised velocity distributions are insensitive to the choice of eccentricity distribution, but the skew and kurtosis are not. The shape of the $v/\vc$ distribution changes considerably as the eccentricity law is varied, but the means and variances are largely unaffected.

There are a number of applications. First, the means and variances are robust against changes in the eccentricity laws and can be used to test against contamination of the sample by unresolved hierarchical systems. Secondly, the discrimination between different eccentricity laws lies in the higher moments of the $v/\vc$ distribution. Using mock data with realistic errors, we demonstrate the viability of our approach, we show how to extract the parameters ($\alpha$ or $\xi$) in our parametrised eccentricity laws using the shape of the normalised velocity distribution. Thirdly, by convolving our analytic distributions with Gaussian uncertainties, we can quantify how much of the tail at $v/\vc \gtrsim \sqrt{2}$ is produced by up-scattering from the errors.


\section*{Acknowledgements}
D.B gratefully acknowledge the supports of the Blavatnik and the Rothschild fellowships. D.B. acknowledges a College Research Associateship at Queens' College, University of Cambridge. We have received partial support from European COST actions CA15117 and CA18108 and the grant KP-06-N38/11. \wyn{We thank Will Sutherland for a helpful referee report.}

\section*{Data Availability} All data available on request to authors.



\bibliographystyle{mnras}
\bibliography{ref}

\begin{thebibliography}{}
\makeatletter
\relax
\def\mn@urlcharsother{\let\do\@makeother \do\$\do\&\do\#\do\^\do\_\do\%\do\~}
\def\mn@doi{\begingroup\mn@urlcharsother \@ifnextchar [ {\mn@doi@}
  {\mn@doi@[]}}
\def\mn@doi@[#1]#2{\def\@tempa{#1}\ifx\@tempa\@empty \href
  {http://dx.doi.org/#2} {doi:#2}\else \href {http://dx.doi.org/#2} {#1}\fi
  \endgroup}
\def\mn@eprint#1#2{\mn@eprint@#1:#2::\@nil}
\def\mn@eprint@arXiv#1{\href {http://arxiv.org/abs/#1} {{\tt arXiv:#1}}}
\def\mn@eprint@dblp#1{\href {http://dblp.uni-trier.de/rec/bibtex/#1.xml}
  {dblp:#1}}
\def\mn@eprint@#1:#2:#3:#4\@nil{\def\@tempa {#1}\def\@tempb {#2}\def\@tempc
  {#3}\ifx \@tempc \@empty \let \@tempc \@tempb \let \@tempb \@tempa \fi \ifx
  \@tempb \@empty \def\@tempb {arXiv}\fi \@ifundefined
  {mn@eprint@\@tempb}{\@tempb:\@tempc}{\expandafter \expandafter \csname
  mn@eprint@\@tempb\endcsname \expandafter{\@tempc}}}

\bibitem[\protect\citeauthoryear{{Ambartsumian}}{{Ambartsumian}}{1937}]{Am37}
{Ambartsumian} V.~A.,  1937, \azh, \href
  {https://ui.adsabs.harvard.edu/abs/1937AZh....14..207A} {14, 207}

\bibitem[\protect\citeauthoryear{Banik}{Banik}{2019}]{1902.01857}
Banik I.,  2019, \mn@doi [Mon. Not. Roy. Astron. Soc.] {10.1093/mnras/stz1551},
  487, 5291

\bibitem[\protect\citeauthoryear{Benisty}{Benisty}{2022}]{Ben22}
Benisty D.,  2022, \mn@doi [Phys. Rev. D] {10.1103/PhysRevD.106.043001}, 106,
  043001

\bibitem[\protect\citeauthoryear{{Brandt}}{{Brandt}}{2018}]{Br18}
{Brandt} T.~D.,  2018, \mn@doi [Astrophys. J.s] {10.3847/1538-4365/aaec06},
  \href {https://ui.adsabs.harvard.edu/abs/2018ApJS..239...31B} {239, 31}

\bibitem[\protect\citeauthoryear{Brax \& Davis}{Brax \& Davis}{2018}]{BD}
Brax P.,  Davis A.-C.,  2018, \mn@doi [Phys. Rev. D]
  {10.1103/PhysRevD.98.063531}, 98, 063531

\bibitem[\protect\citeauthoryear{{Clarke}}{{Clarke}}{2020}]{Cl20}
{Clarke} C.~J.,  2020, \mn@doi [Mon. Not. Roy. Astron. Soc.]
  {10.1093/mnrasl/slz161}, \href
  {https://ui.adsabs.harvard.edu/abs/2020MNRAS.491L..72C} {491, L72}

\bibitem[\protect\citeauthoryear{{El-Badry}}{{El-Badry}}{2019}]{Ba19}
{El-Badry} K.,  2019, \mn@doi [\mnras] {10.1093/mnras/sty3109}, \href
  {https://ui.adsabs.harvard.edu/abs/2019MNRAS.482.5018E} {482, 5018}

\bibitem[\protect\citeauthoryear{{El-Badry} \& {Rix}}{{El-Badry} \&
  {Rix}}{2018}]{ER18}
{El-Badry} K.,  {Rix} H.-W.,  2018, \mn@doi [Mon. Not. Roy. Astron. Soc.]
  {10.1093/mnras/sty2186}, \href
  {https://ui.adsabs.harvard.edu/abs/2018MNRAS.480.4884E} {480, 4884}

\bibitem[\protect\citeauthoryear{{Evans} \& {Oh}}{{Evans} \& {Oh}}{2022}]{EvOh}
{Evans} N.~W.,  {Oh} S.,  2022, \mn@doi [\mnras] {10.1093/mnras/stac707}, \href
  {https://ui.adsabs.harvard.edu/abs/2022MNRAS.512.3846E} {512, 3846}

\bibitem[\protect\citeauthoryear{{Hernandez}, {Jim{\'e}nez}  \&
  {Allen}}{{Hernandez} et~al.}{2012}]{He12}
{Hernandez} X.,  {Jim{\'e}nez} M.~A.,   {Allen} C.,  2012, \mn@doi [European
  Physical Journal C] {10.1140/epjc/s10052-012-1884-6}, \href
  {https://ui.adsabs.harvard.edu/abs/2012EPJC...72.1884H} {72, 1884}

\bibitem[\protect\citeauthoryear{{Hernandez}, {Cort{\'e}s}, {Allen}  \&
  {Scarpa}}{{Hernandez} et~al.}{2019}]{He19}
{Hernandez} X.,  {Cort{\'e}s} R.~A.~M.,  {Allen} C.,   {Scarpa} R.,  2019,
  \mn@doi [International Journal of Modern Physics D]
  {10.1142/S0218271819501013}, \href
  {https://ui.adsabs.harvard.edu/abs/2019IJMPD..2850101H} {28, 1950101}

\bibitem[\protect\citeauthoryear{{Hwang}, {Ting}  \& {Zakamska}}{{Hwang}
  et~al.}{2021}]{HHZ}
{Hwang} H.-C.,  {Ting} Y.-S.,   {Zakamska} N.~L.,  2021, arXiv e-prints, \href
  {https://ui.adsabs.harvard.edu/abs/2021arXiv211101789H} {p. arXiv:2111.01789}

\bibitem[\protect\citeauthoryear{{Jeans}}{{Jeans}}{1919}]{Je19}
{Jeans} J.~H.,  1919, \mn@doi [Mon. Not. Roy. Astron. Soc.]
  {10.1093/mnras/79.6.408}, \href
  {https://ui.adsabs.harvard.edu/abs/1919MNRAS..79..408J} {79, 408}

\bibitem[\protect\citeauthoryear{Juli\'e \& Deruelle}{Juli\'e \&
  Deruelle}{2017}]{FLD}
Juli\'e F.-L.,  Deruelle N.,  2017, \mn@doi [Phys. Rev. D]
  {10.1103/PhysRevD.95.124054}, 95, 124054

\bibitem[\protect\citeauthoryear{{Kervella}, {Arenou}, {Mignard}  \&
  {Th{\'e}venin}}{{Kervella} et~al.}{2019}]{Ke19}
{Kervella} P.,  {Arenou} F.,  {Mignard} F.,   {Th{\'e}venin} F.,  2019, \mn@doi
  [\aap] {10.1051/0004-6361/201834371}, \href
  {https://ui.adsabs.harvard.edu/abs/2019A&A...623A..72K} {623, A72}

\bibitem[\protect\citeauthoryear{{Moe} \& {Di Stefano}}{{Moe} \& {Di
  Stefano}}{2017}]{MS17}
{Moe} M.,  {Di Stefano} R.,  2017, \mn@doi [\apjs] {10.3847/1538-4365/aa6fb6},
  \href {https://ui.adsabs.harvard.edu/abs/2017ApJS..230...15M} {230, 15}

\bibitem[\protect\citeauthoryear{{Murray} \& {Dermott}}{{Murray} \&
  {Dermott}}{1999}]{Mu99}
{Murray} C.~D.,  {Dermott} S.~F.,  1999, {Solar system dynamics}

\bibitem[\protect\citeauthoryear{Pittordis \& Sutherland}{Pittordis \&
  Sutherland}{2018}]{Pi18}
Pittordis C.,  Sutherland W.,  2018, \mn@doi [Mon. Not. Roy. Astron. Soc.]
  {10.1093/mnras/sty1578}, 480, 1778

\bibitem[\protect\citeauthoryear{Pittordis \& Sutherland}{Pittordis \&
  Sutherland}{2019}]{Pi19}
Pittordis C.,  Sutherland W.,  2019, \mn@doi [Mon. Not. Roy. Astron. Soc.]
  {10.1093/mnras/stz1898}, 488, 4740

\bibitem[\protect\citeauthoryear{{Pittordis} \& {Sutherland}}{{Pittordis} \&
  {Sutherland}}{2022}]{Pi22}
{Pittordis} C.,  {Sutherland} W.,  2022, arXiv e-prints, \href
  {https://ui.adsabs.harvard.edu/abs/2022arXiv220502846P} {p. arXiv:2205.02846}

\bibitem[\protect\citeauthoryear{{Tokovinin}}{{Tokovinin}}{1998}]{To98}
{Tokovinin} A.~A.,  1998, Astronomy Letters, \href
  {https://ui.adsabs.harvard.edu/abs/1998AstL...24..178T} {24, 178}

\bibitem[\protect\citeauthoryear{{Tokovinin}}{{Tokovinin}}{2020}]{To20}
{Tokovinin} A.,  2020, \mn@doi [Mon. Not. Roy. Astron. Soc.]
  {10.1093/mnras/staa1639}, \href
  {https://ui.adsabs.harvard.edu/abs/2020MNRAS.496..987T} {496, 987}

\bibitem[\protect\citeauthoryear{{Tokovinin} \& {Kiyaeva}}{{Tokovinin} \&
  {Kiyaeva}}{2016}]{To16}
{Tokovinin} A.,  {Kiyaeva} O.,  2016, \mn@doi [Mon. Not. Roy. Astron. Soc.]
  {10.1093/mnras/stv2825}, \href
  {https://ui.adsabs.harvard.edu/abs/2016MNRAS.456.2070T} {456, 2070}

\bibitem[\protect\citeauthoryear{{Zahn}}{{Zahn}}{1977}]{Za77}
{Zahn} J.~P.,  1977, Astron. Astrophys., \href
  {https://ui.adsabs.harvard.edu/abs/1977A&A....57..383Z} {500, 121}

\bibitem[\protect\citeauthoryear{{Zhao}, {Li}  \& {Bienaym{\'e}}}{{Zhao}
  et~al.}{2010}]{ZBB}
{Zhao} H.,  {Li} B.,   {Bienaym{\'e}} O.,  2010, \mn@doi [\prd]
  {10.1103/PhysRevD.82.103001}, \href
  {https://ui.adsabs.harvard.edu/abs/2010PhRvD..82j3001Z} {82, 103001}

\makeatother
\end{thebibliography}

\end{document}